%% file: HSPBSPmaster.tex
\documentclass[journal]{IEEEtran}
\usepackage{graphicx} 
\usepackage{subcaption}
\usepackage{amsmath}
\usepackage{float}
\usepackage{mathtools}
\usepackage{amssymb}
\usepackage[font=footnotesize,labelfont=bf]{caption}
\usepackage{tabularx}
\usepackage{textcomp}
\usepackage{soul}
\usepackage{color}
\usepackage{booktabs}
\usepackage{tikz}

\makeatletter
\def\maketag@@@#1{\hbox{\m@th\normalfont\normalsize#1}}
\makeatother

\newcommand\copyrighttext{%
  \footnotesize \textcopyright 2021 IEEE. Personal use of this material is permitted. Permission from IEEE must be obtained for all other uses, in any current or future media, including reprinting/republishing this material for advertising or promotional purposes, creating new collective works, for resale or redistribution to servers or lists, or reuse of any copyrighted component of this work in other works.}
\newcommand\copyrightnotice{%
\begin{tikzpicture}[remember picture,overlay]
\node[anchor=south,yshift=10pt] at (current page.south) {\fbox{\parbox{\dimexpr\textwidth-\fboxsep-\fboxrule\relax}{\copyrighttext}}};
\end{tikzpicture}%
}

\begin{document}
\bstctlcite{IEEEexample:BSTcontrol} \title{Impulse data models for the inverse problem of electrocardiography}

\author{Tommy Peng, Avinash Malik, Laura R. Bear, and Mark L. Trew \thanks{Tommy Peng is with the Department of Electrical and Computer Engineering, University of Auckland. E-mail: tpen280@aucklanduni.ac.nz.}\thanks{Avinash Malik is with the Department of Electrical and Computer Engineering, University of Auckland.}\thanks{Laura Bear is with IHU-LIRYC.}\thanks{Mark L. Trew is with the Auckland Bioengineering Institute, University of Auckland.}\thanks{Manuscript submitted March 31, 2021. This work was funded by the National Science Challenges, Science for Technological Innovation, NZ, grant 3713917. Mark L. Trew was supported by a grant from the Fondation Leducq.}}
\maketitle
\copyrightnotice

\input{abstract.tex}
\input{introduction.tex}
\input{methods_restructured}
\input{results_restructured}
\input{discussion.tex}
\input{conclusion.tex}
\bibliographystyle{IEEEtran}
\bibliography{HSPBSPmaster}
\end{document}

%% file: abstract.tex
\begin{abstract}
  \textit{Objective:} To develop, train and test neural networks for predicting heart surface potentials (HSPs) from body surface potentials (BSPs). The method re-frames traditional inverse problems of electrocardiography into regression problems, constraining the solution space by decomposing signals with multidimensional Gaussian impulse basis functions. \textit{Methods:} Impulse HSPs were generated with single Gaussian basis functions at discrete heart surface locations and projected to corresponding BSPs using a volume conductor torso model. Both BSP (inputs) and HSP (outputs) were mapped to regular 2D surface meshes and used to train a neural network. Predictive capabilities of the network were tested with unseen synthetic and experimental data. \textit{Results:} A dense full connected single hidden layer neural network was trained to map body surface impulses to heart surface Gaussian basis functions for reconstructing HSP. Synthetic pulses moving across the heart surface were predicted from the neural network with root mean squared error of $9.1\pm1.4$\%. Predicted signals were robust to noise up to 20 dB and errors due to displacement and rotation of the heart within the torso were bounded and predictable. A shift of the heart 40 mm toward the spine resulted in a 4\% increase in signal feature localization error. The set of training impulse function data could be reduced, and prediction error remained bounded. Recorded HSPs from in-vitro pig hearts were reliably decomposed using space-time Gaussian basis functions. Activation times calculated from predicted HSPs for left-ventricular pacing had a mean absolute error of $10.4\pm11.4$ ms. Other pacing scenarios were analyzed with similar success. \textit{Conclusion:} Impulses from Gaussian basis functions are potentially an effective and robust way to train simple neural network data models for reconstructing HSPs from decomposed BSPs. \textit{Significance:} The HSPs predicted by the neural network can be used to generate activation maps that non-invasively identify features of cardiac electrical dysfunction and can guide subsequent treatment options.
\end{abstract}



%% file: introduction.tex
\section{Introduction}
\label{sec:introductionNN}

Predicting heart surface potentials (HSPs) from body surface potentials (BSPs) is known as the inverse problem of electrocardiography \cite{Messinger-Rapport1988}. Good predictions enable non-invasive heart surface analysis which can assist source localization for cardiac anti-arrhythmic procedures such as ablation~\cite{Zemzemi2013}. The forward relationship between HSPs and BSPs is described in Eq~\ref{eq:forwarddescription}, where $A$ characterizes the geometric relationship between the body and heart surface potential at any instance in time. The inverse problem of estimating $A^{-1}$ given instances of concurrent BSPs and HSPs is known to be ill-posed and underdetermined~\cite{Messinger-Rapport1988, Pullan2010}. Therefore the inverse problem is difficult to solve. Current approaches such as electrocardiographic imaging (ECGI) \cite{Ghosh2009, Yoram1999} estimate the HSP ($\hat{HSP}_{\lambda}$) by minimizing the error or L2-norm residual ($||A \times HSP - BSP||^{2}_{2}$) with an additional penalty term $\lambda R$ (Eq~\ref{eq:regdescription}), where $\lambda$ is a specialized regularization parameter and $R$ is the regularizor. A large regularization parameter results in overly smooth HSP predictions, and a parameter too small causes oscillating predictions \cite{Messinger-Rapport1988, Brooks1999}. ECGI resolves this issue by using a number of electrophysiological constraints and special smoothing operators~\cite{Yoram1999}. While methods exist for estimating the quasi-optimal regularization parameter, there is no single technique which performs best for all geometries and signal-to-noise ratios~\cite{Colli-Franzone1985a, Colli-Franzone1985, Pullan2010}. Nevertheless, ECGI is a well documented technique for solving the cardiac inverse problem\mbox{~\cite{Shahidi1994, VanDam2009}}, and comprehensive reviews of methods and validation  can be found in\mbox{~\cite{Pullan2010, Milanic2014, Karoui2018}}.

\begin{equation}
  BSP = A \times HSP
  \label{eq:forwarddescription}
\end{equation}
\begin{equation}
  \hat{HSP}_{\lambda} = argmin(||A \times HSP - BSP||^{2}_{2} + \lambda^{2} R)
  \label{eq:regdescription}
\end{equation}

Predicting HSPs from BSPs can also be solved via unsupervised learning regression data driven approaches, which model the BSP-HSP relationship by using simultaneously recorded BSP-HSP pairs. Examples of regression solutions include clustered support vector machine\mbox{~\cite{Jiang2012}}, time-delayed neural networks\mbox{~\cite{Karoui2019, Malik2018}}, relevance vector regression~\mbox{\cite{Giffard-Roisin2019}}, and auto-encoder\mbox{~\cite{Ghimire2019, Bujnarowski, Bacoyannis2021}} methods. These are different from data driven regularization techniques\mbox{~\cite{Ding2019, Cluitmans2018}}, which explicitly rely on a regularization parameter within the model. A common challenge encountered by data driven models is generating a sufficiently rich set of data to form a comprehensive training set~\cite{Bujnarowski, Karoui2019}. This difficulty arises when data models are trained or fitted using full BSPs and full HSPs. Trained data driven models can only predict data similar to those found in the training set, so generalizing a regression solution for the BSP-HSP problem requires many paired BSP and HSP recordings across widely different heart states. However, detailed simultaneous HSP and BSP recordings from hearts spanning all possible states is difficult to obtain. Previous studies have augmented data using simulations of \textit{full} HSPs and BSPs\mbox{~\cite{Ghimire2019, Karoui2019, Giffard-Roisin2019}}. In this work, it is \textit{hypothesized} that the neural network solution for the cardiac inverse problem can be learned using \textit{impulse} basis functions that span the space of HSPs under different heart conditions. Recorded potential signals can then be decomposed into the same impulse basis functions and each component predicted separately and then combined to reconstruct the complex signal.

To test this hypothesis, HSP and BSP impulse basis function pairs are generated with a volume conductor model in a heart-torso domain~\cite{Bear2015}. Gaussian impulse basis functions have been shown to be effective for decomposing cardiac signals~\cite{McSharry2003a}. They have been used to model signals under different drug and disease conditions~\cite{Badilini2008, Peng2018, Peng2019}. Unlike orthonormal basis functions such as sinusoidal or wavelets which often have multiple peaks or troughs, the energy in Gaussian functions is concentrated in a single peak with bound support, which makes it suitable for physiological interpretation\mbox{~\cite{Roonizi2013a}}.

In this work, we develop an approach for predicting heart surface potentials using signal decomposition, signal projection, and neural networks. \textbf{Our contributions} are: (1) showing that HSPs can be effectively modelled and generated using Gaussian 3D basis functions; (2) demonstrating that a single neural network trained on Gaussian basis functions can be used to predict unseen physiologically relevant synthesized HSPs from BSPs; (3) performing prediction of HSPs from experimental heart recordings. The proposed method is significant as it generalizes the BSP-HSP regression problem by learning in the Gaussian function space which can represent electrocardiographic signals from various disease and drug states. Our approach predicts the output of a system by learning the system characteristics through basis function driven responses.


%% file: methods_restructured.tex
\section{Methods}
\label{sec:methodsNN}
\begin{figure*}[bth]
  \centering
  \includegraphics[width=0.60\textwidth]{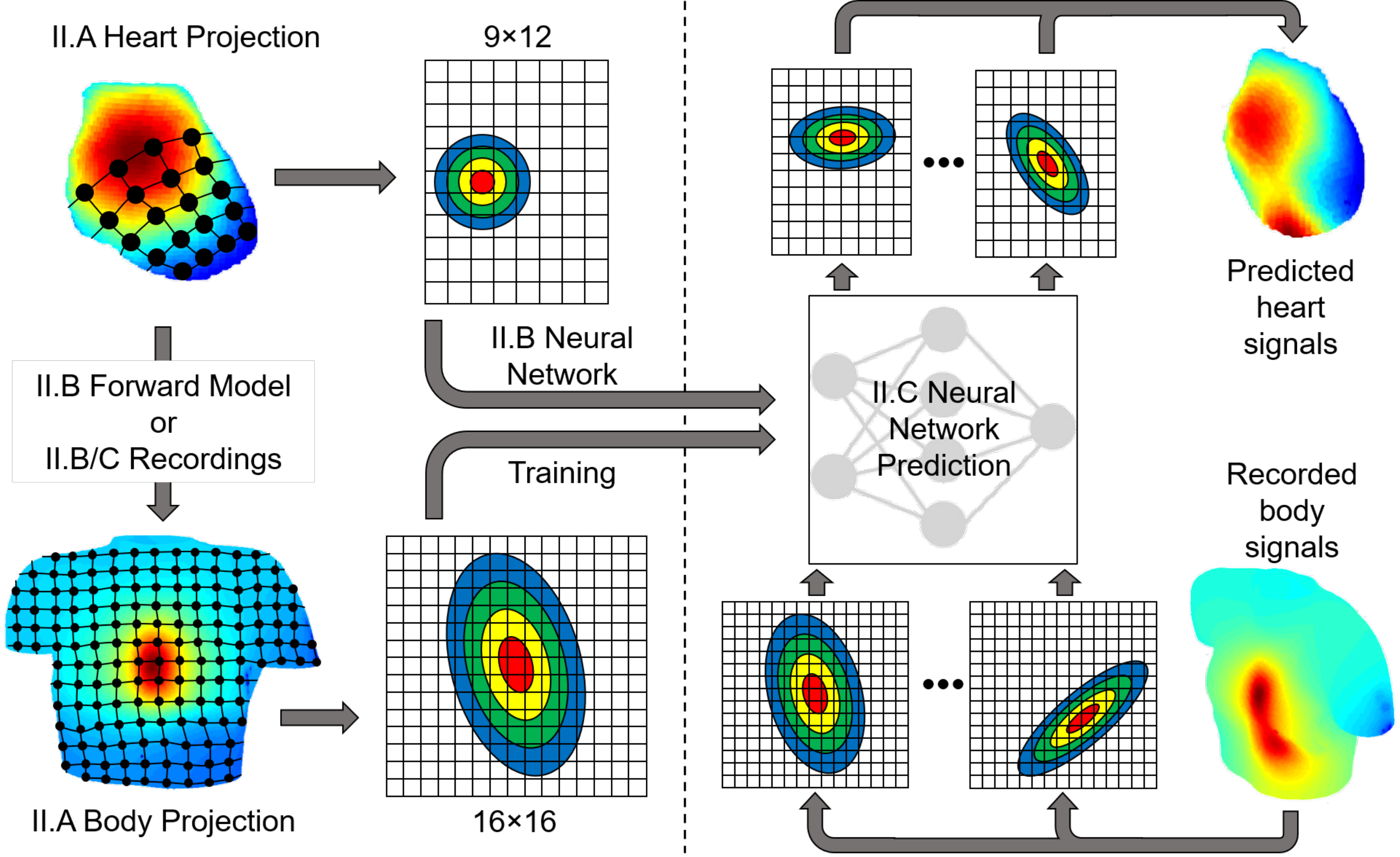}
  \caption{Neural network training and prediction pipeline.}
  \label{fig:nnoverall}
\end{figure*}

Figure~\ref{fig:nnoverall} is a visual representation of the Gaussian basis neural network pipeline. The forward model is used for generating training and testing data. It can be replaced or augmented with  experimental data when sufficient of these are available.

\subsection{Data Context}
\label{sec:methodsNN:datapreNNdes}
Heart and body surfaces were abstracted to regular 2D meshes~\cite{Le2018}. These provided a standard context for training and applying neural network data models. They are also consistent with the topology of electrode matrices used for electrical mapping on both the heart and body surfaces~\cite{Bear2018}. Consequently, the heart surface was represented by a mesh of 9-by-12 points and the body surface a mesh of 16-by-16 points. The body surface mesh dimensions were specified as powers of 2, with the total numbers of points just greater than the typical numbers of body surface electrodes.

The arrays of heart surface points were obtained directly from the “sock” of recording electrodes used experimentally~\cite{Bear2018} and the body surface points from electrodes embedded in a torso tank. The body surface points were projected onto a cylindrical torso approximation and the cylinder was then unwrapped via a cut along the projected left anterior descending artery line identified in the heart surface data. The location of the artery line was projected to the cylinder in the same way as the body surface points. The unwrapped cylinder surface was then re-sampled onto a 16-by-16 body surface mesh using linear interpolation. See Supplementary Materials for more details on the 2D abstractions.

\subsection{Neural Network Data Model}
\label{sec:methodsNN:NNtrain}

\subsubsection{Neural Network Design and Implementation}
\label{sec:methodsNN:NNtrain:1}
This work shows that a neural network approach can capture the basic relationship between HSP impulses and their respective BSPs. To that end, a simple neural network with one dense fully connected hidden layer architecture was implemented in Python using Keras with a Tensorflow back-end\mbox{~\cite{chollet2015keras, tensorflow2015-whitepaper}}. This common network architecture is mathematically proven to be capable of approximating any smooth continuous function given enough neurons and appropriate activation functions~\cite{Cybenko1989, Funahashi1989}. The network was designed to capture the relationship in the $x$-$y$ plane between 2D Gaussian basis HSPs and the simultaneous 2D BSPs. Therefore, the input layer consisted of 256 neurons corresponding to the 16-by-16 body surface 2D abstraction points, and the output layer consisted of 108 neurons corresponding to the 9-by-12 heart surface electrodes. Neuron bias was turned off, as we expected a zero BSP input to produce a zero HSP output. 

\subsubsection{Training Data Generation}
\label{sec:methodsNN:NNtrain:2}
Gaussian functions have been successfully used to decompose cardiac signals into linear sums of bases\cite{McSharry2003a, Peng2018, Badilini2008}. Decomposition enables signals from different states to be expressed in a common parameter space so that objective representation and quantitative comparisons can be made. Other studies have used orthonormal basis functions, such as wavelets\mbox{~\cite{Ding2019, Cluitmans2018}}, for efficient representation of electrocardiographic signals. However, projecting multiphasic wavelet signals from heart to body surface through the effectively low pass torso filter causes loss of discriminatory information such as signal peaks. In contrast, we used monophasic Gaussians to generate a feature rich training set. A Gaussian basis function at a integer point ($x$, $y$) on the 9-by-12 heart surface mesh (Fig~\ref{fig:nnoverall}) is:

\begin{multline}
G(x, y) = \\
\mathcal{A} \times exp(-((x-\mu_{x})^2+(y-\mu_{y})^2)/(2\sigma^2))
\label{eq:2dgaussian}
\end{multline}

The Gaussian width parameter, $\sigma$, is the same in both $x$ and $y$. The position ($\mu_{x}$, $\mu_{y}$) is the center of the Gaussian and $\mathcal{A}$ is the amplitude. A library of training impulses was constructed by enumerating $\sigma$ in steps of 0.02 between 0.01 and 0.5 (of full mesh size). For each combination of $\mu_{x}$, $\mu_{y}$ and $\sigma$ values, the amplitudes ($\mathcal{A}$) a vary in value according to a half Gaussian function with peak value of +1 or -1 (to give both positive and negative impulses for training) and a Gaussian width of 0.15 sampled at 24 equispaced intervals. A width of 0.15 ensured a variety of amplitudes between -1 and 1 were selected. In total 129600 distinct Gaussian impulses were generated on the heart surface mesh. 

Each heart surface impulse was projected to the body surface using a volume conductor torso model~\cite{Bear2015}. The model was based on the realistic geometry from a porcine heart in a human shaped torso tank experimental set up\mbox{~\cite{Bear2018}}. The specific forward model is not a critical component of this work and could be substituted by other models or experimental simultaneous and comprehensive heart and body surface recordings if these are available. The experimental recordings would need to be for impulse stimuli or the impulse components of the stimuli separable from the remainder of the heart surface signals. The 129600 HSP-BSP pairs were used as the neural network training and testing data set.

\subsubsection{Training Regime}
\label{sec:methodsNN:NNtrain:3}
A 70-30 training-validation split was applied to the data pairs. A 0.1 dropout rate was used to reduce over-fitting. The hyperparameters for the neural net were tuned through grid search of possible parameter pairings. The grid search was performed in Python with \textit{GridSearchCV} from scikit-learn. The hyperparameter search space is shown in \mbox{Table~\ref{tab:hyper}}, with the best hyperparameters highlighted in green. Training was done using the Adaptive Momentum Estimation (ADAM) optimizer with root mean squared loss function. Training was completed over 100 epochs.

\begin{center}
\captionof{table}{Hyperparameter Search Space} 
\footnotesize
\begin{tabular}{ |c|c| } 
\toprule
 Hidden Neurons & [100, 150, 200, 250, 300, 350, \textcolor{green}{400}]\\
 Activation Function & [ReLU, Sigmoid, \textcolor{green}{TanH}]\\ 
 Regularization & [\textcolor{green}{None}, L1, L2]\\
 Batch Size & [16, 32, \textcolor{green}{64}]\\
 Learning Rate & [0.1, 0.01, \textcolor{green}{0.001}]\\
 \bottomrule
\end{tabular}
 \label{tab:hyper}
\end{center}

\subsection{Predictions using Parameterized Neural Networks}
\label{sec:methodsNN:NNpred}
Predictions used data sets \textit{not} seen by the neural network during training. During prediction, the inputs to the neural network are the 2D BSPs, and the outputs from the neural network are the 2D HSPs. All predicted stationary HSPs were evaluated using error, which was calculated per 2D potential surface ($x$-$y$ plane) as percent root mean squared error (RMSE) (Eq~\ref{eq:nrmse}), where $\hat{p}$ is the predicted by the neural network and $p$ is the the Gaussian specified by Eq~\ref{eq:2dgaussian} for that data pair.

\begin{equation}
  RMSE(\hat{p}, p) = 100\dfrac{\sqrt{mean((\hat{p} - p)^{2})}}{max(p) - min(p)}
  \label{eq:nrmse}
\end{equation}

Input and predicted Gaussians were also compared using Euclidean distances between signal peaks (peak Euclidean distance) measured on the HSP mesh. Predicted experimental HSPs were compared using activation time, which was defined to be the time point for the most negative signal gradient (dV/dt).

\subsubsection{Pure Moving Gaussians}
\label{sec:methodsNN:NNpred:1}
\begin{figure}[bth]
  \includegraphics[width=\columnwidth]{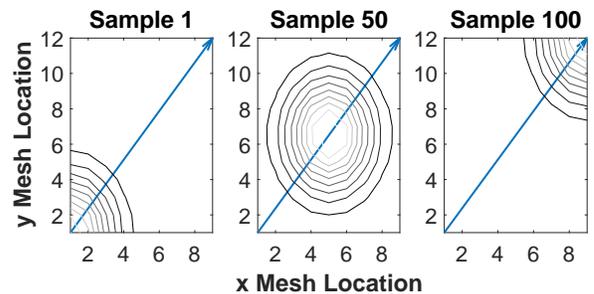}
  \caption{Synthesized moving Gaussian data Path 1 (from point [1, 1] to [9, 12]) with contours of the moving Gaussian.}
  \label{fig:pathOnSurfContour}
\end{figure}

A physiologically relevant testing set was generated comprising signals moving across the 2D heart surface array. Gaussian functions of fixed amplitude and standard deviation were displaced along straight line paths between two points in the 9-by-12 array (p1 and p2 in Table~\ref{tab:test2points}). The Gaussians were evaluated with their peaks at 100 discrete points along each path. Fig~\ref{fig:pathOnSurfContour} shows an example for three positions along Path 1. In this example $\sigma$ was set to 0.177. The six paths were chosen to abstractly reflect common cardiac pacing locations and activation paths. Each HSP Gaussian in the testing set was passed through the volume conductor torso model to generate 600 BSP-HSP pairs for testing. The BSPs were used as input for the neural network and used to predict 2D HSPs.

\begin{center}
\captionof{table}{$x$, $y$ Dimension Location of Test I Points} 
\footnotesize
\begin{tabular}{ |c|c|c|c|c|c|c|c| } 
\toprule
 Path & 1 & 2 & 3 & 4 & 5 & 6\\
 \midrule
 p1 & [1, 1] & [1, 6] & [1, 12] & [4.5, 1] & [4.5, 12] & [9, 1]\\ 
 p2 & [9, 12] & [9, 1] & [1, 6] & [9, 12] & [9, 1] & [9, 12]\\
 \bottomrule
\end{tabular}
 \label{tab:test2points}
\end{center}

\begin{figure}[bth]
  \includegraphics[width=\columnwidth]{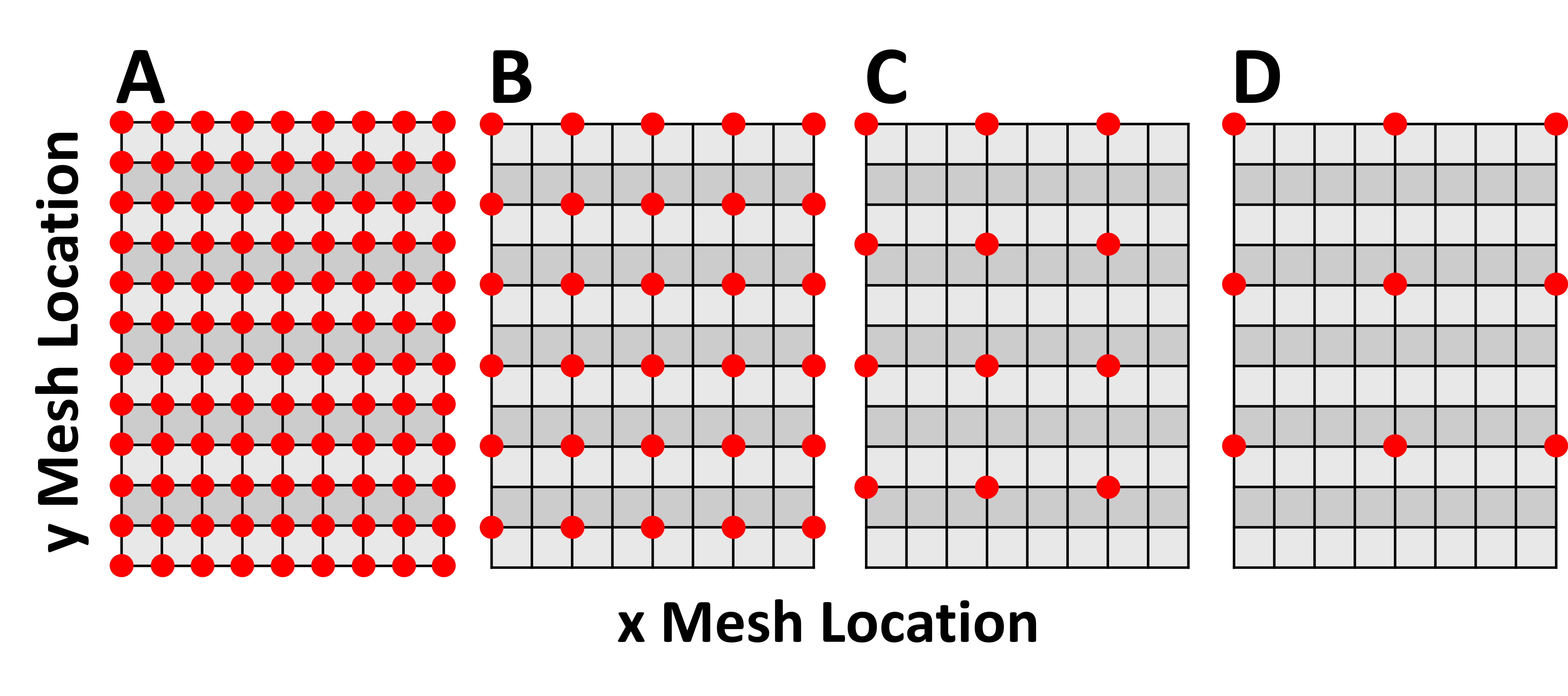}
  \caption{The locations of training set peak ($\mu$) locations in the $x$-$y$ (9-by-12) plane of the HSP surface. Red dots indicate the points in the $x$-$y$ plane with a training set peak. A) original training set from Section~\ref{sec:methodsNN:NNtrain:1}; B) reduction scheme 1; C) reduction scheme 2; D) reduction scheme 3.}
  \label{fig:downsamplePeakLocationScheme}
\end{figure}

Additionally, moving Gaussian signals were used to explore the impacts of three modifications to training data on predicted HSPs:
\begin{enumerate}
    \item \textbf{Noise Robustness Test}: Recorded signals are often noisy, and this contributes to poor predictions from neural networks. Using similar approaches to previous studies~\cite{Karoui2019}, BSPs generated from moving Gaussian signal HSPs were augmented with Gaussian white noise (MATLAB\textregistered~\textit{awgn} function) at 2dB, 5dB, 10dB, 20dB, and 50dB signal-to-noise ratio. This dataset enabled quantification of the trained neural net sensitivity to noise in the BSP signal.
    \item \textbf{Geometry Robustness Test}: Between subjects, the heart location varies relative to the torso and body surface. We simulated these variations and/or body geometry measurement errors by: (1) translating the heart position toward the spine in steps of 5 mm, up to 40 mm; (2) rotating the heart around its long axis in $10^{\circ}$ increments between $-40^{\circ}$ and $40^{\circ}$. The rotation axis was determined using principle component analysis on the heart electrode locations. A positive rotation is clockwise around the axis pointing toward the base of the heart. These translations and rotations were performed in 3D to modify the heart location. The torso tank geometry was unchanged throughout. The subsequent altered forward models were then used to produce BSPs from HSPs along all 6 paths of the moving Gaussian testing set. The BSPs were used to predict HSPs using the neural network trained for the original heart position.
    \item \textbf{Reduced Training Set Test}: Reductions in the training set were used to explore how this would affect the prediction results. An effective training data set for neural networks should include examples from all expected prediction outcomes~\cite{Thirumalainambi2003}. We applied 3 reduction schemes for the $x$-$y$ dimension peak location as shown in subplots B, C, D of Fig~\ref{fig:downsamplePeakLocationScheme}. We split the original G3D training set into 25 different $\sigma$ reduction schemes consisting of training sets with a single $x$-$y$ dimension $\sigma$ at 0.02 intervals between 0.01 and 0.5 for HSPs. Each of the reduced training sets were used to train a separate neural network.
\end{enumerate}

The prediction results from the three data modification tests were compared against the pure moving Gaussian predictions from the neural net trained on the full training set. 

\subsubsection{Experimental Recorded Data}
\label{sec:methodsNN:NNpred:2}
All procedures were approved by the ethical committee of Bordeaux CEEA50 and adhered to Directive 2010/63/EU of the European Parliament on protection of animals used for scientific purposes. A single excised porcine heart was Langendorff perfused within a human torso shaped tank and an epicardial sock with 108 electrodes was placed around the ventricles~\cite{Bear2018}. The tank contained 128 embedded surface electrodes and was filled with saline solution (conductivity 0.3 mS/mm). Simultaneous epicardium and body surface electrical potential recordings were made for sinus rhythm. Afterwards, left bundle branch block was induced via lesions produced by ablation. Left, right, and bi-ventricular pacing recordings were subsequently recorded\mbox{~\cite{Bear2018}}. The geometry of the heart in the tank was extracted from CT images. This geometry was used to construct the volume conductor forward model for this study and as the basis of ECGI comparative predictions\mbox{~\cite{Bear2018}}.

We used averaged beat recordings from the experimental data which are time 649 samples in length, recorded at a sampling rate of 2048Hz~\cite{Bear2018}. This resulted in an LV HSP recording matrix of 9-by-12-by-649 after 2D abstraction at each sample in time.

The 2D spatial Gaussian of Eq~\ref{eq:2dgaussian} was extended to include time as:

\begin{equation}
      G3D(x, y, t) = \mathcal{A} \times G(x, y) \times exp(\dfrac{(t - \mu_{t})^{2}}{2 \sigma_{t}^{2}}))
  \label{eq:g3dbasis}
\end{equation}

The HSP 9-by-12-by-649 array was determined from fitted G3D basis functions using an extension of generalized orthogonal forward regression~\cite{Badilini2008, Peng2019}. Potential values were normalized between -1 and 1 while maintaining relative scale and sign. To fit each G3D component, the unexplained signal was first correlated against a library of G3D basis functions. A description of the library can be found in the Supplementary Materials. Then, the library function with the largest absolute correlation value was set as the initial point for loss function minimization using MATLAB\textregistered~function \textit{fmincon}. The loss function employed was the sum of squared differences (Eq~\ref{eq:sumsqdiff}), where $v$ is the 1D form of the 3D matrix to be fitted and $\hat{v}$ is the 1D form of the current fit of the 3D matrix. The bounds of optimization are shown in Table~\ref{tab:fminconbounds}. 

\begin{equation}
  SSD(v, \hat{v}) = \sum (v - \hat{v})^{2}
  \label{eq:sumsqdiff}
\end{equation}

\begin{center}
\captionof{table}{Table of fmincon Optimization Bounds} 
\begin{tabular}{ |c|c|c|c|c|c|c|c| } 
\toprule
 Bound & $\mu_{x}$ & $\mu_{y}$ & $\mu_{t}$ & $\mathcal{A}$ & $\sigma$ & $\sigma_{t}$ \\ 
 \midrule
 Upper & 9 & 12 & 1 & Max pot. & 0.5 & 0.5\\ 
 Lower & 1 & 1 & 0 & Min pot. & 0 & 0\\
 \bottomrule
\end{tabular}
 \label{tab:fminconbounds}
\end{center}

The bounds on $\mu$ ensure the peaks of the G3D components were bounded within the recording, the $\mathcal{A}$ bounds allow the amplitude of the G3D peak to vary between the range of potentials within the recording, the $\sigma$ and $\sigma_{t}$ bounds enable fitted G3D components to have visible effect across the whole heartbeat. The bounds here are similar to those of~\cite{Peng2019}, which have been shown as effective for describing electrocardiographic signals. The fitted G3D component was subtracted from all basis functions in the G3D library leaving an unexplained signal to be fitted in the subsequent iteration. The process iterated until the number of predefined G3D components were fit to the 3D recording matrix. The final G3D fit of the recording was evaluated per 2D potential surface ($x$-$y$ plane) using $RMSE$ where the potential range is 2 from the normalization of potential between -1 and 1.

The HSP from a left ventricular averaged heartbeat was fitted with 100 G3D basis functions. Each basis function was split into the 649 time instances, and each time instance was an HSP boundary condition to the volume conductor torso model to simulate corresponding BSPs. This created a testing set of 64900 2D surface BSPs that were used as inputs for the neural network to back-predict HSPs for comparison with experimental recordings.

%% file: results_restructured.tex
\section{Results}
\label{sec:resultsNN}

\subsection{Prediction of Moving Pure Gaussian HSPs}
\begin{figure}[tbh]
\centering
  \includegraphics[width=0.8\columnwidth]{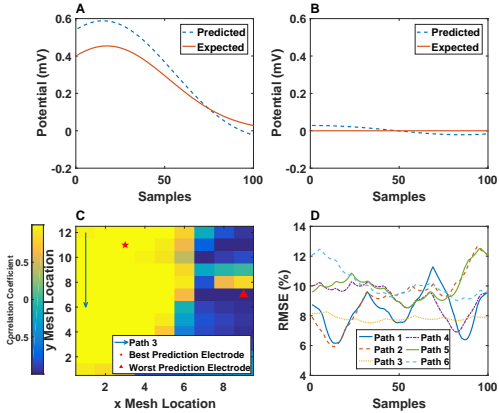}
  \caption{Moving Gaussian predictions. A) Time series prediction results for best prediction correlation at electrode (x = 3, y = 11) for Path 3 in Table~\ref{tab:test2points}. B) Time series prediction results for worst prediction correlation at electrode (x = 9, y = 7) for Path 3 in Table~\ref{tab:test2points}. C) Path 3 on the heart surface 9-by-12 grid, along with the time series correlation between predicted and synthetic data at each electrode. The electrode with the worst prediction in correlation is marked with a red triangle, while the electrode with best prediction in correlation is marked with a red star. D) RMSE for the predicted 2D sample slices (along x-y axis) for all 6 paths of synthesized HSPs moving from point to point.}
  \label{fig:predGen_ts}
\end{figure}

A neural network trained solely with Gaussian impulses on discrete HSP and BSP grids reliably reconstructed continuous Gaussian signals moving across the heart surface grid. An example of time series comparison for Path 3 can be seen in Fig~\ref{fig:predGen_ts}A-C. Fig~\ref{fig:predGen_ts}D shows the RMSE between original and predicted moving Gaussians on the heart surface. Across the six paths described in Table~\ref{tab:test2points}, the RMSE is $9.12\pm1.37$\%. 

\subsection{Robustness Tests}
\begin{figure}[tbh]
\centering
  \includegraphics[width=0.9\columnwidth]{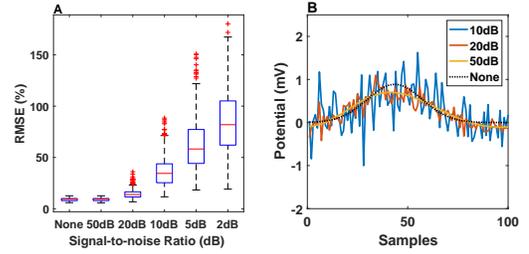}
  \caption{Effects of noise added to BSPs. A) Boxplot of RMSE value across 2D HSP slices ($x$-$y$ plane) between predicted and expected HSPs at different noise levels for the input BSP. B) Time series example of predicted HSPs at different BSP noise levels for path 1 from Table~\ref{tab:test2points}. The expected prediction is the synthesized HSPs moving from point to point (dotted line).}
  \label{fig:noiseCombo}
\end{figure}

The moving Gaussian BSP was augmented with five levels of white noise (Section~\ref{sec:methodsNN:NNpred:1}) and each signal used as input to the neural network. Fig~\ref{fig:noiseCombo}A shows distributions of RMSE between raw HSP and predicted HSP signals from noisy BSP. As expected, the error associated with the prediction increases as the noise level increases. Fig~\ref{fig:noiseCombo}B gives an example of the HSP time series prediction for none, 40, 20 and 10 dB added noise.
 
\begin{figure}[tbh]
\centering
  \includegraphics[width=0.8\columnwidth]{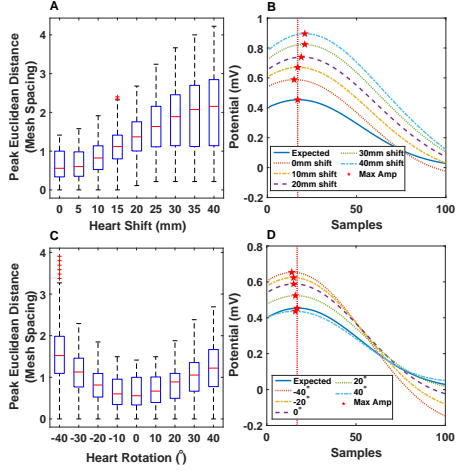}
  \caption{Effect of moving and rotating heart positions on neural network predictions. A) Boxplot of Euclidean distance between predicted and expected peaks of HSPs at different spatial shifts for the heart geometry in the forward model. B) Time series example of predicted HSPs at different shifts in heart geometry for Path 3 from Table~\ref{tab:test2points}. C) Boxplot of Euclidean distance between predicted and expected peaks of HSPs at different rotation angles for the heart geometry in the forward model. D) Time series example of predicted HSPs at different shifts in heart geometry for Path 3 from Table~\ref{tab:test2points}. In B and D, the expected peak timing is marked as the red dotted vertical line.}
  \label{fig:noiseRotateHeartShift}
\end{figure}

The location of the heart relative to the torso was translated and rotated in discrete steps and BSPs were generated for moving Gaussians on the heart surface. The altered BSPs were used as inputs to the neural network trained solely with Gaussian impulses on the originally positioned heart. Fig~\ref{fig:noiseRotateHeartShift}A (translation) and Fig~\ref{fig:noiseRotateHeartShift}C (rotation) show that error grows in a predictable and bounded way. For a sizeable shift of 40 mm toward the spine, the peak of the predicted Gaussian is a median distance of around 1.6 mm further from the raw data position than no shift (i.e. ~4\% increase in error for a 40 mm shift). Fig~\ref{fig:noiseRotateHeartShift}B shows that for translation the predicted signal morphology is similar to signals from the training position but with increasing amplitude and with the peak offset from the raw data. Fig~\ref{fig:noiseRotateHeartShift}D indicates that a negative rotation (left ventricle rotates toward the front of the chest) increases amplitude and decreases width compared to no-rotation, whereas a positive rotation (right ventricle rotates toward the front of the chest) has the opposite effect.

\subsection{Effects of Reduced Training Sets}
\begin{figure*}[tbh]
\centering
  \includegraphics[width=0.6\textwidth]{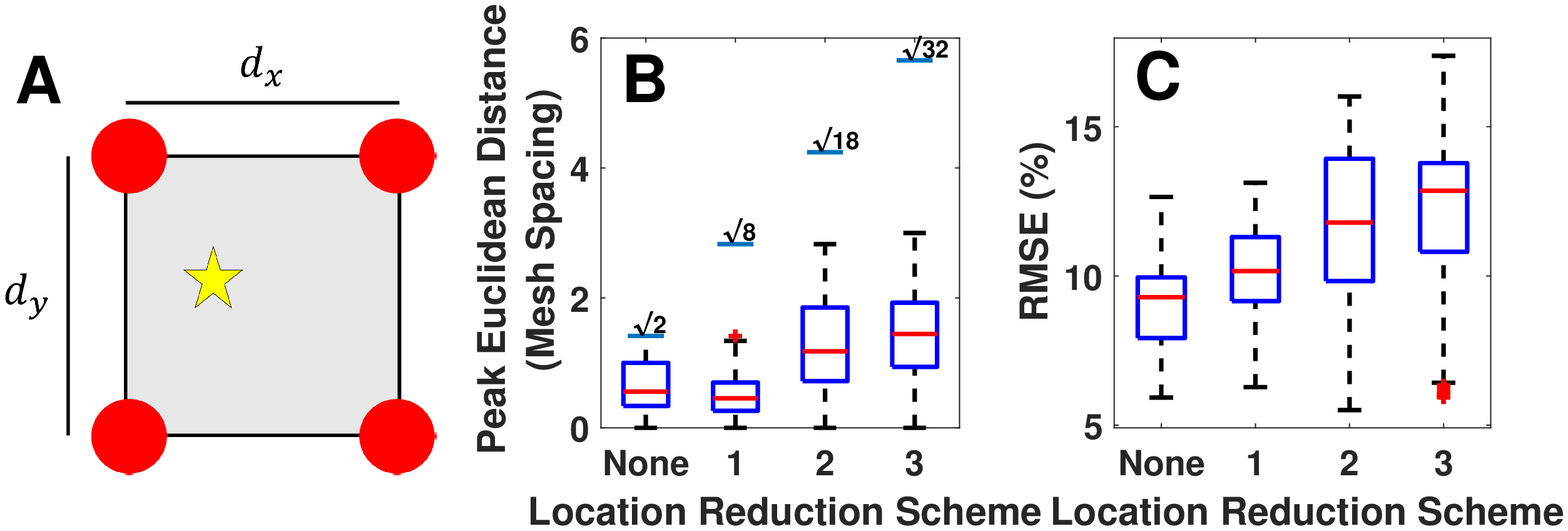}
  \caption{Effects of reduced size training data sets. A) A small section of the HSP electrode grid. Electrodes are shown as red circles and expected peak of the predicted HSP is shown as the yellow star. The distance between electrodes is $d_{x}$ and $d_{y}$ in the x and y direction, respectively. B) Euclidean distance in the $x$-$y$ plane between predicted and expected Gaussian peak locations for the synthetic moving Gaussian testing data set. The proposed limits on maximum Euclidean distance between predicted and expected Gaussian peaks is shown as the blue line for each box plot. C) RMSE per 2D slice.}
  \label{fig:downsampleboxplotpeak}
\end{figure*}

Fig~\ref{fig:downsampleboxplotpeak}A shows a subsection of a heart surface electrode grid formed by four neighboring electrodes. For training the neural network, Gaussian impulses with peaks at the electrode points were used to generate equivalent body surface signals using a volume conductor forward torso model. The training data (or electrode) reduction tests (Fig~\ref{fig:downsamplePeakLocationScheme}) increase the distance between impulse peaks ($d_{x}$ and $d_{y}$). To predict an HSP Gaussian with a peak bounded by the reduced electrode training set the maximum Euclidean distance error for predicted peak location is bounded by: $\sqrt{d^{2}_{x}+d^{2}_{y}}$.

Fig~\ref{fig:downsampleboxplotpeak}B shows that Euclidian distance errors for reduced training data are bounded as expected. Fig~\ref{fig:downsampleboxplotpeak}C shows that the RMSE increases as the number of examples found in the training set decreases.

\begin{figure}[tbh]
\centering
  \includegraphics[width=0.9\columnwidth]{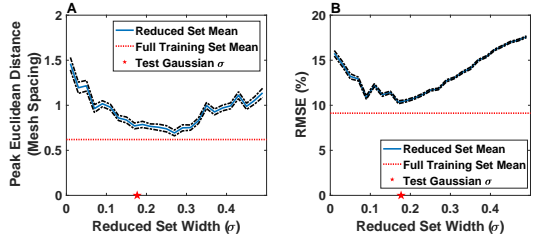}
  \caption{The mean of the errors when predicted using neural networks trained with $\sigma$ reduction scheme sets is shown as the blue solid line, the 95\% confidence intervals are marked by the black dotted lines. The $\sigma$ of the testing set Gaussian is shown as the red star. For comparison, the mean of the errors when the full training set is used is the red dotted line. A) Euclidean distance in the $x$-$y$ plane between predicted and expected Gaussian peak locations for the synthetic moving Gaussian testing data set; B) RMSE per 2D slice.}
  \label{fig:tieredwidthpeak}
\end{figure}

Neural networks were trained using 25 different Gaussian width reduction sets, each training set included Gaussians of a single chosen width ($\sigma$ from Eq~\ref{eq:2dgaussian}) at 0.02 intervals between 0.01 and 0.5. Fig~\ref{fig:tieredwidthpeak}A shows mean and standard deviation changes of the Euclidian peak distance over the altered widths, for predicted and expected signals. Fig~\ref{fig:tieredwidthpeak}B shows mean and standard deviation of RMSE over the altered widths, for predicted and expected signals. The standard deviation $\sigma$ for the moving Gaussian in the testing set is 0.177 (red star). In Fig~\ref{fig:tieredwidthpeak}, both Euclidian distance and RMSE are minimized as $\sigma$ for the training data approaches 0.1768. 

\subsection{G3D Decomposition of Recorded HSPs}
\begin{figure}[tbh]
\centering
  \includegraphics[width=\columnwidth]{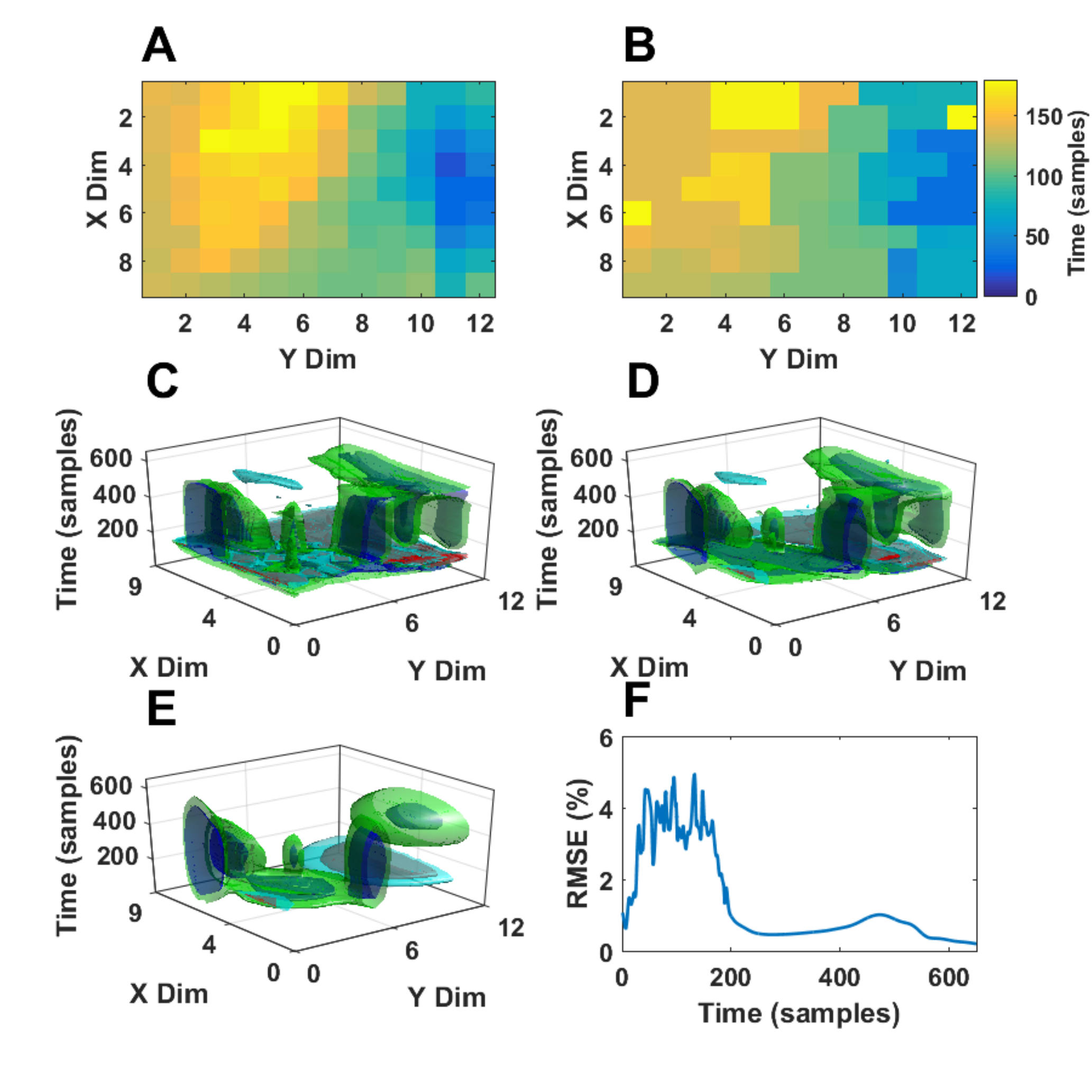}
  \caption{Comparisons between G3D representation of real-world recordings. A) The activation map corresponding to LV recordings. B) The activation map from the G3D representation. C) Isosurfaces for the full 9-by-12-by-649 matrix for the left ventricular recording. D) Isosurfaces for the G3D representation. E) Isosurfaces for the sum of the first 10 fitted G3D components. All isosurfaces are drawn at values of -0.1, -0.05, 0.05, and 0.1 mV to show the various potentials found within the 3D signal matrices. F) RMSE per 2D slice ($x$-$y$ plane) for the G3D representation.}
  \label{fig:isosurfaces_combo}
\end{figure}

The 3D space-time matrix of size 9-by-12-by-649 for the left ventricular recording was fitted using 100 G3D components. Original signals were approximated by summing the 100 components. Figs~\ref{fig:isosurfaces_combo}A and ~\ref{fig:isosurfaces_combo}B compare activation times extracted from the raw experimental signal with those extracted from the G3D signal model. The potential isosurfaces around 0 mV in space and time are contrasted between the raw data and the 100 component G3D model in Figs~\ref{fig:isosurfaces_combo}C and~\ref{fig:isosurfaces_combo}D. The development of the approximation is shown in Fig~\ref{fig:isosurfaces_combo}E where 10\% of the G3D components are used to show the model captures principal space-time features. For the full reconstruction of HSP on the heart surface grid, the peak RMSE is less than 5\% (Fig~\ref{fig:isosurfaces_combo}F), with mean RMSE of $1.34\pm1.30$\%. 

\subsection{Prediction of Recorded HSPs}
\begin{figure*}[tbh]
\centering
  \includegraphics[width=0.80\textwidth]{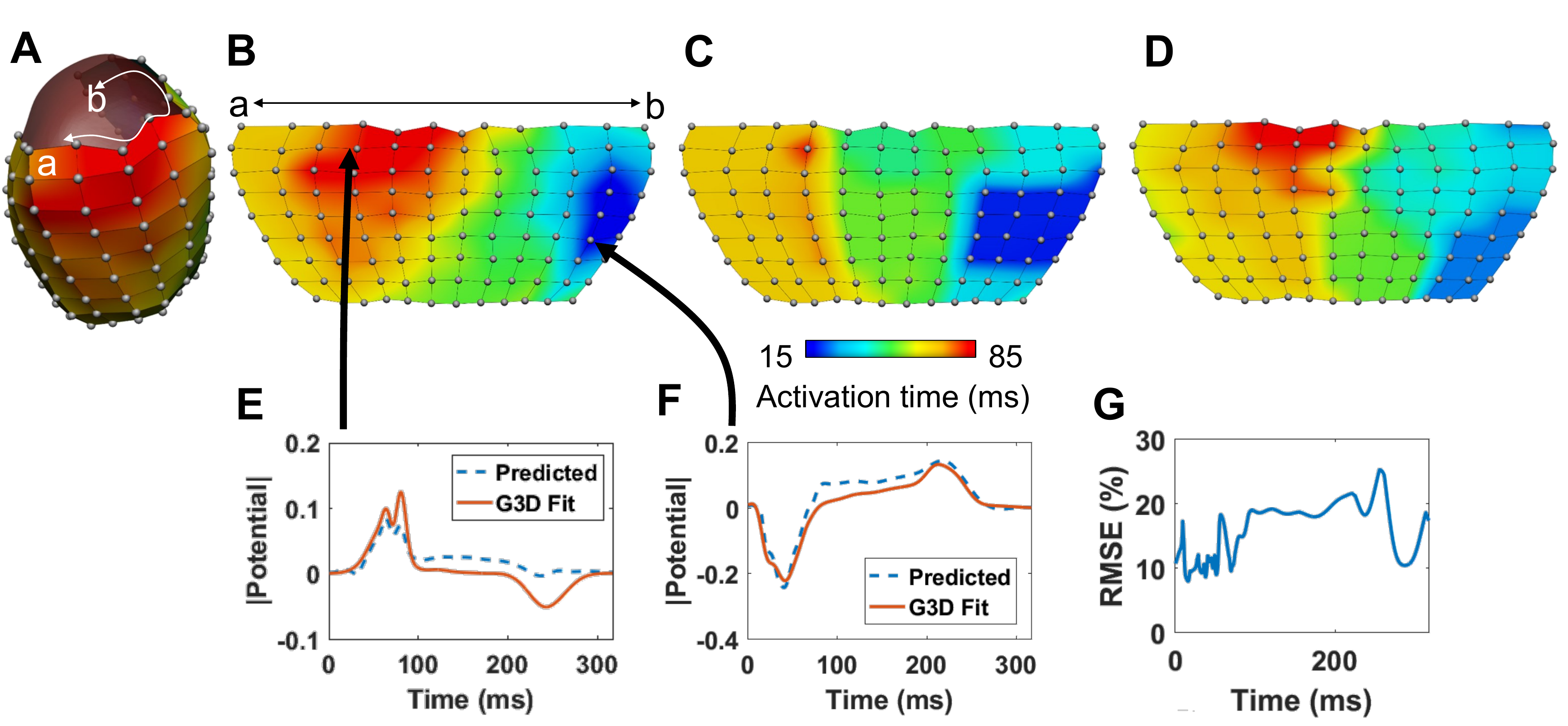}
  \caption{Comparisons between recorded and predicted heart surface activation and normalized signals. A) Recording electrode sock in place on heart. B) Recorded activation sequence for left-ventricular pacing. C) Impulse predicted activation sequence for left-ventricular pacing. D) ECGI activation sequence for left-ventricular pacing from~\mbox{\cite{Bear2018}}. E) and F) Comparison between recorded fitted normalized signals and predicted signals for indicated points. G) Global root mean squared error for 2D potential surfaces across time.}
  \label{fig:realworldCombo}
\end{figure*}

In Fig~\ref{fig:realworldCombo}, left ventricular paced activation times derived from experimental recorded signals are compared with neural network predicted HSPs and ECGI predicted HSPs across the ventricular epicardial electrode sock. The absolute difference in calculated activation times from experimental recordings and neural network predictions, across all electrodes, was $10.4\pm11.4$ms for impulse prediction and $8.10\pm7.17$ms for ECGI. The RMSE in predicted activation times is 17.7\% for impulse prediction and 12.4\% for ECGI. Fig~\ref{fig:realworldCombo}E and~\ref{fig:realworldCombo}F compare predicted HSP signals with G3D fits of the corresponding recording. Fig~\ref{fig:realworldCombo}G summarizes the spatial RMSE between HSP predicted by the neural network and the G3D description of raw recorded data following left-ventricular pacing of an in-vitro heart~\cite{Bear2018}. The mean RMSE is $16.5\pm4.18$\%. Peak differences occur during the plateau and repolarization phases of the activation. Using the same techniques and neural network as left-ventricular pacing, predictions of activation sequences for sinus rhythm, bi-ventricular pacing and right-ventricular pacing were compared with experimental recordings. The corresponding activation maps for these scenarios can be found in the Supplementary Materials.

%% file: discussion.tex
\section{Discussion}
\label{sec:discussion}
In this paper, we present the concept for an impulse basis function neural network approach to solving the inverse problem of electrocardiography. We have shown that a neural network can be used to model the data relationships between components of electrical signals on the body and heart surfaces, for a torso model. We trained the neural network with static Gaussian bases and used it to predict synthetic signals and recorded heart surface signals. The proposed method is significantly different from traditional neural network models which learn the relationship between \textit{full} HSP and BSPs. In the context of our problem the method is robust to noise and perturbed heart locations. This is an important first step toward developing a new approach to the inverse problem of electrocardiography. 

The Gaussian basis function impulses on the heart surface along with simultaneous forward model calculated data on the body surface were used to train a neural network which predicted both synthesized HSPs and recorded decomposed HSPs from BSPs. The network performs well for predicting the synthesized moving Gaussian data set described by Table~\ref{tab:test2points}, with low mean RMSE value of 8.46\% across all 6 paths. As the neural net was only trained using basis functions with peaks at integer $x$ and $y$ mesh locations, the prediction error and Euclidean distance are expected to fluctuate (Fig~\ref{fig:predGen_ts}D) along all 6 paths which include peaks at non-integer $x$ and $y$ dimensions. Interestingly, Fig~\ref{fig:predGen_ts}A suggests that the time series at electrode locations that are closer to the points of activation (i.e. along the path) achieve better prediction results when compared to electrodes which are far from the point of activation. This is most likely due to the distant electrodes having near zero values in time series (Fig~\ref{fig:predGen_ts}B), which is difficult to predict for an neural net largely trained on impulse or non zero time series data.

One of the advantages of the proposed data driven model is that it learns the relationship between impulse HSPs and their respective BSPs. Previously we showed that Gaussian impulses with finite parameter bounds can effectively represent electrocardiographic signals from a variety of drug and disease states\mbox{~\cite{Peng2018, Peng2019}}. An impulse data driven model trained on appropriate data within these finite bounds can predict HSPs under a variety of heart conditions through linear combination of predicted Gaussian impulses. We show that the proposed method predict HSPs under different pacing conditions, before and after ablation induced left bundle branch block, but the technique is equally applicable to other heart states, if they are not already in the training set. To predict HSPs from subjects under new heart states, the neural network training set would need to be augmented with Gaussian impulses which are similar to those decomposed from the new heart states. This is a fundamental property of predictive data driven models, as they can only predict what it has seen before in the training set.

The trained neural network is susceptible to noise in the input BSPs as seen in Fig~\ref{fig:noiseCombo}. There is a large drop in predicted HSP quality when going from BSP inputs with 20dB to 10dB signal-to-noise ratio in Fig~\ref{fig:noiseCombo}A. Furthermore, Fig~\ref{fig:noiseCombo}B suggests that, noisier BSPs will produce noisier HSPs as neural network outputs. Similar prediction error trends are observed by spatial adaptation time delayed neural networks from\mbox{~\cite{Karoui2019}}. This is expected as the network is trained on data without noise. Appropriate BSP pre-processing can improve prediction of HSPs for ECGI~\cite{Bear2021}. It is feasible to consider similar techniques for BSP noise reduction for the proposed neural network approach. 

The proposed impulse response method addresses data driven model generalization across different cardiac rhythms. In the same way inter-subject geometric variability can be sampled and encoded in the data models. \mbox{Fig~\ref{fig:noiseRotateHeartShift}A} shows that the neural network prediction is worse for hearts that are increasingly distant from the spatial location of the heart used to generate the training set. The time series in \mbox{Fig~\ref{fig:noiseRotateHeartShift}B} shows that the shapes of the predicted HSP signal are largely unchanged between 0mm and 40mm heart location shifts. Furthermore, in \mbox{Fig~\ref{fig:noiseRotateHeartShift}}, there is a non-random change in predicted signal amplitude, peak time, and width as the heart shifts and rotates. Ongoing work looks to extend heart and torso projections into common reference frames~\mbox{~\cite{Giffard-Roisin2019}}. The common reference projections implicitly encode geometry. One way to augment this is to modify neural network architecture and explicitly include geometry measures as model inputs.

We observe several important factors for constructing an effective training set for prediction of Gaussian basis HSPs using neural nets. We evaluated the effectiveness of training sets based their effect on the Euclidean distance between predicted and expected Gaussian peaks. This is an important measure, as most interesting heart behaviors occur when the electrical potential on the heart surface is non-zero. We show that the maximum Euclidean distance error between predicted and expected Gaussian peaks on the heart surface is related to the distance between Gaussian peaks found in the training set. Specifically, if the training set has examples of peaks at distance $d_{x}$ in the $x$ dimension of the heart surface, and $d_{y}$ in the $y$ dimension of the heart surface, then the maximum error in Euclidean distance of the predicted peak is $\sqrt{d^{2}_{x}+d^{2}_{y}}$. This limit is made for the proposed model under the assumption that the training set HSP-BSP pairs at neighbouring electrodes are sufficiently different, and the difference between training set HSP-BSP pairs between neighbours' neighbours are larger than for neighbours alone. In Fig~\ref{fig:tieredwidthpeak} we show that the neural network trained on the full training set outperforms any of the neural nets trained on reduced sets with G3D basis functions having a single $\sigma$. Therefore, for our proposed data driven model, an effective training set must include a fair representation of data from varying $\sigma$ values. This is similar to the guidelines for effective training sets proposed by~\cite{Thirumalainambi2003}.

We have shown that a G3D basis model can effective capture the morphological behavior found within HSPs expressed as a 3D matrix ($x$, $y$, $t$). The G3D basis model offers descriptions of the HSP 3D matrix which improves with the number of G3D components fit (Supplementary Materials). This is common behavior for basis function models~\cite{Roonizi2013a}. Due to the smooth nature of the Gaussian basis, its representation of a given recording is good but not exact, as seen in Fig~\ref{fig:isosurfaces_combo}. This is similar to the behavior noted in~\cite{Peng2019}, where small components electrocardiographic signals were sometimes not captured by the Gaussian model. The G3D basis function decomposition technique proposed here is not signal dependent, and can decompose, and therefore express, HSPs from varying heart states in the same basis function space. The optimization bounds found in \mbox{Table~\ref{tab:fminconbounds}} and loss function can be modified based on domain knowledge for effective representation of different heart states\mbox{~\cite{Peng2019, Peng2018}}.

The neural network prediction of G3D decomposed HSP left ventricular recordings is notable, as the neural network is trained on the Gaussian basis function HSPs and forward model BSPs solutions, and \textbf{not} with a data set of simultaneously recorded BSPs and HSPs. The proposed impulse trained model predicts real-world HSP data with an error of 16.5\% \mbox{(Fig~\ref{fig:realworldCombo}G)}, which is similar to prediction errors found using other neural networks trained on full real-world HSP and BSP data~\mbox{\cite{Bujnarowski, Karoui2019}}. The predicted activation map \mbox{(Fig~\ref{fig:realworldCombo}C)} are similar to those found via ECGI \mbox{(Fig~\ref{fig:realworldCombo}D)}. The dynamic behavior of the recorded HSP is predicted well. This can be seen in the volatile activation within the first 200 samples, the relatively flat period between 200 and 400 samples, and the repolarization period after 400 samples in which the potential returns to baseline for both the recorded and predicted time series in Figs~\ref{fig:realworldCombo}E-G. This is further supported by the mean absolute difference in predicted and expected activation time of 10.7 ms.

We present the prediction of real-world signals based on neural networks trained with basis function data to serve as a case study for how such an approach can be used to solve clinical problems. Source localization accuracy is an important factor to establish before the use of HSP prediction techniques in clinical settings. For ECGI, in vivo validation studies\mbox{~\cite{Ghanem2001, Bear2018}} require significant organization, certification, and authorization to carry out. While the proposed Gaussian impulse data trained neural network model can predict experimentally recorded HSPs and activation maps, validation studies must be done to establish its localization accuracy before its use in clinical settings. To that end, the forward model used to generate the training set could be replaced by measured data, or limited measured data can be augmented by a model. While developing this method, a model-based BSP generator reduces some uncertainties, and furthermore, impulse data from heart and body surface is currently not widely available. Currently, stimulus artifacts from pacing experiments provide limited data\mbox{~\cite{Hooks2007}}. In the future, it can be hypothesized that these impulse experimental recordings can be made by applying Gaussian impulse patterns on the surface of a model heart, with realistic conductivity and geometries, similar to the manikin human head used to record impulse response characterizations in the audiology domain\mbox{~\cite{Chen1995}}.

Unsurprisingly, we find that the quality of predictions increases as the number of HSP-BSP pairs in the training set increases. In particular, we show that there is a maximum error in Euclidean distance of predicted Gaussian peaks. This limit can be used to calculate the target distance between stimulus and recording electrodes for experimental HSP-BSP recordings to train our approach and produce HSP predictions at a desired error level. Based on successful decomposition of HSPs into Gaussian components, ongoing research uses similar curve fitting techniques\mbox{~\cite{Peng2019, Peng2018}} to decompose BSP into components similar to those in the training data set. A robust fitting solution to decompose BSPs is an important step towards clinical adoption.


%% file: conclusion.tex
\section{Conclusion}
In this work we propose a Gaussian basis function decomposition approach to bypass the rich training data problem of neural network models for the inverse problem of electrocardiography. We have shown that a network trained on basis function HSPs and their respective generated BSPs can be used to predict experimental recordings of HSPs decomposed into the same basis function set. The proposed data driven model can also be trained using experimentally recorded HSP impulse and their respective BSPs.